# DEMAND ENGINEERING: IP NETWORK OPTIMISATION THROUGH INTELLIGENT DEMAND PLACEMENT

JOHN EVANS (CISCO), ARASH AFRAKTEH (CISCO), RUOYANG XIU (GOOGLE)

## 1. ABSTRACT

Traffic engineering has been used in IP / MPLS networks for a number of years as a tool for making more efficient use of capacity by explicitly routing traffic demands where there is available network capacity that would otherwise be unused.   Deployment of traffic engineering imposes an additional layer of complexity to network design and operations, however, which has constrained its adoption for capacity optimisation.   The rise of software defined networks (SDN) has renewed interest in the use of traffic engineering approaches leveraging centralised network controllers for capacity optimisation.

We argue that future networks can realise the network optimisation benefits of traffic engineering without incurring additional network complexity through closer coupling between the network and the applications and services using the network.   This can be achieved through leveraging a network- and traffic-aware controller to directly influence where applications and services site or locate service instances, i.e. which implicitly impacts the paths that the application's or service's traffic demands take through the network.   We call this technique Demand Engineering.   Demand Engineering has the additional benefit of providing an admission control capability, i.e. which can provide an assurance that network SLAs can be met.

In this paper we describe the concept of Demand Engineering, give examples of its use and present simulation results indicating its potential benefits.   We also compare demand engineering to traffic engineering.

## 2. BACKGROUND

The Cisco Visual Networking Index (VNI) predicts that global Internet traffic will grow 3.0-fold from 2015 to 2020 [1], a compound annual growth rate of 25%.   In the presence of such traffic growth, network operators are faced with the sometimes competing challenges of providing network Service Level Agreement (SLA) assurances whilst managing their costs by making efficient use of network capacity and minimising operational expenses. To achieve this, ideally their network topology and installed capacity would be a close match to their network traffic demands, i.e. through effective network engineering and capacity planning processes they ensure that capacity is installed where it is needed.   In practise, however, constraints and costs in underlying networks can result in suboptimal network topologies which together with inaccuracies in planning and unexpected traffic demands can result in an imbalance between the installed capacity and the actual traffic demand load.   This imbalance creates risks either of not meeting committed SLAs due to insufficient provisioned bandwidth or of bandwidth over-provisioning resulting in inefficient use of capacity and unnecessary cost. Such risks are exacerbated in ever more dynamic service environments, where services can be created or terminated on demand, with rapidly changing traffic profiles and service churn. Network operators currently have a limited toolset to deal with these problems.

Traffic engineering (TE) – either through manipulation of routing protocol metrics [2][3] or through MPLS-based RSVP-TE [4] – provides a possible solution to these problems. TE attempts to optimise for short term differences between the installed capacity and the offered traffic load by routing traffic demands where there is available network capacity which would otherwise be unused; the result is an effective bandwidth gain. TE has been employed in IP / MPLS networks for more than 10 years, however, deployment of TE imposes an additional layer of complexity to network design and operations and until recently only a minority of service providers actively used TE for capacity optimisation[1] [5]. The rise of software defined networks (SDN) and centralised network controllers have renewed interest in the use of traffic engineering approaches for capacity optimisation [7] [8].

Traffic engineering is commonly deployed as a network-layer capability with no direct feedback to the higher layer applications and services that use the network. Traffic engineering can be augmented by admission control mechanisms such as RSVP [9] to provide closed loop feedback, i.e. providing the capability to verify whether there is sufficient capacity to support the SLA required by a new application or service prior to it being deployed, and to signal back to the application or service if not. In practise, however, RSVP has failed to gain widespread use for admission control.

Today, most network operators address the problems of SLA assurance and capacity management through over-provisioning and offering loose SLA assurances.

[10] and [11] show the potential benefits of closer coupling between the network and the applications and services using the network in meeting application SLAs and making efficient use of capacity, specifically in the case of peer-to-peer (P2P) overlay applications. Through providing limited network topology information to P2P applications they were able to demonstrate both a reduction in network utilisation and an improvement in application performance.

In this paper we present demand engineering which is a more generalised solution that is able to address the problems of providing network SLA assurances whilst making efficient use of network capacity through close coupling between the network and the applications and services using the network.

## 3. DEMAND ENGINEERING OVERVIEW

We define Demand Engineering as the process whereby a requesting application or service needs to site or locate an application or service instance which will require a set of network resources; the application/service requests these resources from the Demand Engineering Controller which replies with the "best" location to site the service. "Best" is determined using an understanding of the network topology and traffic and is defined in terms of both being able to meet the SLA requirements and making effective use of the network capacity.

---

[1] RSVP-TE has other uses as an IP/MPLS network-layer signalling protocol and to provide fast reroute (FRR) capabilities [6].

Demand engineering is applicable to applications and services where there is a choice as to where the application endpoints may be located, e.g.:

- *Cloud-based xAAS offerings:* which is the best data centre (DC) to site an Infrastructure-as-a-Service (IAAS), Platform-as-a-Service (PAAS) or software-as-a-Service (SAAS) workload?
- *Network Function Virtualisation:* where is the best location to site a virtualised network function (VNF)?
- *Content delivery networks:* which is the best cache to use to serve a particular piece of content?
- *Web applications:* which is the best application server to use for a particular request?
- *Peer-to-peer overlays:* which is the best source peer for a particular piece of content?
- *Wireless backhaul-optimisation:* which access point or base station for a client end point to use to best make use of backhaul capacity?

Considered in the context of the Cisco VNI predictions, these uses cases potentially address ~90% of the traffic generated by applications and services in 2016.

Demand engineering addresses the dual problem spaces covered by traffic engineering and admission control. Demand engineering directly influences the location of traffic sources and destinations, i.e. which implicitly impacts the paths that the application or service's traffic demands take through the network. Through understanding the impact that requested demands will have on the network the demand engineering controller is able to determine which locations are acceptable and which is best. Demand engineering closes the feedback loop between the network and the applications and services that use the network.

## 4. DEMAND ENGINEERING PRINCIPLES

We use the example of placement of a cloud-based xAAS workload[2] to describe the principles of demand engineering in more detail; consider the system that implements the demand engineering process shown in Figure 1.

---

[2] A workload consists of compute, storage, and network resources; in this example we consider the network resources only.

Figure 1.    Demand engineering system

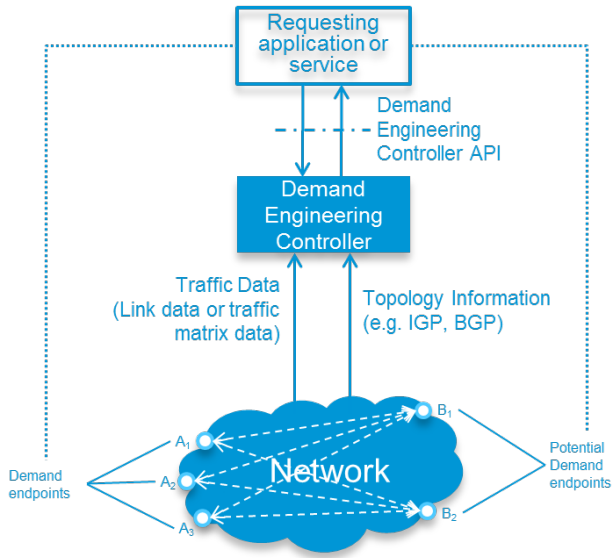

1.  *Requestor.*    A requesting application or service needs to create a new xAAS workload; the workload will impose new bandwidth demands on the network.    There are a set of potential DC locations where the workload may be sited {$B_1$, $B_2$, … $B_n$}.

2.  *Demand Engineering Controller API.*    Prior to creating the new workload, the requestor seeks guidance from a controller via an API on which of the set of potential locations to use.

    This request may include the following parameters:

    a)  The A-end addresses for the traffic demands, i.e. which are fixed and may represent the set of locations of application instances that will be accessing the workload: {$A_1$, $A_2$, …, $A_n$}

    b)  The B-end addresses for the traffic demands, i.e. which represent the set of DC locations where the workload may be sited: {$B_1$, $B_2$, … $B_n$}

    c)  The bandwidth required; this may be asymmetrical and may differ between traffic demand legs: $D_{A1 \rightarrow Bx}$, $D_{A2 \rightarrow Bx}$, … $D_{Bx \rightarrow A1}$, $D_{Bx \rightarrow A2}$, …

    d)  The maximum acceptable latency: $L_{max}$

    e)  The network failure sets of interest, i.e. in which network failure cases the service needs to be guaranteed, e.g. no failures, link failures, node failures, shared risk failures.

3.  *Demand Engineering Controller.*    The controller is network- and traffic-aware and is able to determine what impact the new demands will have on the network and to determine which location is best, where best means is able to meet the network SLA requirements and is optimal as determined by an optimisation policy.

    This requires that the demand engineering controller has a view of the network topology, the network traffic demand matrix (which may be measured or deduced), and the network routing behaviour (e.g. which may use an IGP, BGP, RSVP-TE etc.), such that it can determine the traffic utilisation of each link in the network with the new demands added through network simulation,

taking into account failure cases if necessary.    This view of the network can extend end-to-end across both Layer 2 and Layer 3.

The controller may determine the traffic demand matrix from the network traffic state (which may be the current state or from a previous time period) or from reservation state (which may be for current or future reservations), or from a combination of both approaches.

4. *Demand Engineer Controller Decision Logic.*    The decision made by the demand engineering controller may be considered as an extension to an admission control decision, e.g.:

    a) *Admission control:* considering a specific A, B site pair:

        i. What is the minimum capacity remaining (R) on all of the links on the paths between site $A_1$ and site $B_1$ (taking into account failure cases if necessary) if demand D is added?

            1. R < (1-T): is an admission control failure, which would indicate that there is insufficient bandwidth to support the requested SLA because the network utilisation exceeds the threshold of maximum acceptable bandwidth utilisation (T).    T is set to ensure that the loss and jitter SLAs required for the workload can be met.    An admission control failure indication would be returned to the requestor.

            2. R > (1-T): there is sufficient bandwidth ➔ consider latency

        ii. What is the maximum latency ($L_{pmax}$) on all of the paths between site $A_1$ and site $B_1$ (similarly taking into account failure cases)?

            1. $L_{pmax} > L_{max}$: is an admission control failure, which would indicate the latency is too high to support the requested SLA because the path latency exceeds the maximum acceptable latency ($L_{max}$).    An admission control failure indication would be returned to the requestor.

            2. $L_{pmax} < L_{max}$: admission control success, i.e. indicating that the SLA can be met ➔ apply optimisation policy.

This process is repeated for all A, B pair combinations; the set of combinations for which there has been an admission control success determines the list of feasible sites

It is also possible to consider other factors such as network availability and loss in the admission control decision.

    b) *Optimisation:* if there are multiple feasible sites apply an optimisation policy to determine the preferred location.

# 5. DEMAND ENGINEERING OPTIMISATION OBJECTIVE

Demand engineering optimisation can be defined as an optimisation problem where the objective is to find the best of all possible solutions. Given a fixed network topology, an existing set of traffic demands and new demands to place, the optimisation problem can be defined as determining the placement of those new demands that satisfies the SLA requirements and best meets the optimisation objective. In order to solve this problem, we need to define the optimisation objective.

A number of optimisation objectives are possible. It is possible to optimise the network SLA, e.g. by minimising latency. For many applications and services, however, if admission control has already determined that the required latency can be met there is limited benefit in further reducing the network latency. Hence we focus on defining an optimisation objective which will make efficient use of network capacity.

The underlying aim of making efficient use of network capacity is to minimise costs. It is potentially possible to define an explicit optimisation objective to minimise the network cost for an individual demand engineering request. Minimising the cost of an individual request, however, does not assure that the overall costs across a set of requests is minimised. Hence we focus on defining an optimisation objective which will make efficient use of network capacity overall as a proxy to minimise the network capacity costs across a set of requests.

We first consider the *network worst-case utilisation* which is a common metric used in traffic engineering optimisation to make efficient use of installed capacity. Networks are subject to component failures hence it is normal to provision sufficient capacity to deal with the traffic rerouting that will occur under such failure events. *Worst-case utilisation* can be used to characterise the network utilisation in a set of potential failure scenarios, e.g. no failures, single circuit failures, single node failures, shared risk failures; for a particular link, the worst-case utilisation under a set of failure scenarios is the maximum utilisation determined for that link over all of the failure scenarios in the set. Worst-case link utilisations are useful for identifying bottlenecks in the network that will only become apparent when a particular failure occurs. The overall *network worst-case utilisation* is the maximum worst-case link utilisation over all links in the network under evaluation, which summarises how resilient the network is to network failure from a capacity perspective. Network worst-case utilisation is inversely related to the maximum resilient throughput achievable in a network; this is the maximum throughput achievable, assuming that all demands grow uniformly, before the utilisation on any link, in any failure scenario of interest, exceeds a defined threshold. If the network worst-case utilisation is minimised, the achievable resilient throughput is maximised, i.e. the net effect is that you can support more traffic on the same network before you have to upgrade capacity. Hence, minimisation of the network worst-case utilisation is a simple goal that is often used to guide traffic engineering optimisation.

In demand engineering, however, new demands may only impact a subset of the network topology, and therefore network worst-case utilisation may be determined by parts of the network which are not impacted by the new demands. Hence network worst-case utilisation is not specific enough to use as an optimisation goal; instead we focus on the *worst-case path utilisation* to guide optimisation through demand engineering. The *worst-case path utilisation* is the maximum worst-case link utilisation over all links on all of the network paths that would be used by the new demands, taking into account the failure scenarios of interest. Hence for demand engineering minimising the worst-case path utilisation for each new set of demands serves to maximise achievable throughput.

# 6. DEMAND ENGINEERING EXAMPLE

To explain the practise of demand engineering, we add detail to the previous example and consider a specific case of cloud-based xAAS workload placement.

Consider the network shown in Figure 2, which has:

- 4 candidate DC sites {chi, nyc, kcy, sjc} – ringed in red, i.e. these represent the potential B-end locations.
- 11 access sites {chi, nyc, kcy, sjc, lax, sea, hst, atl, mia, wdc, bos} – shown as blue squares; these provide access connections to remote branch sites and represent the sources or destinations of traffic demands to/from the DCs, i.e. they represent the set of A-end locations.
- IGP-based forwarding with link metrics proportional to circuit latency, i.e. such that demands will follow the lowest latency routes

Figure 2. Example network

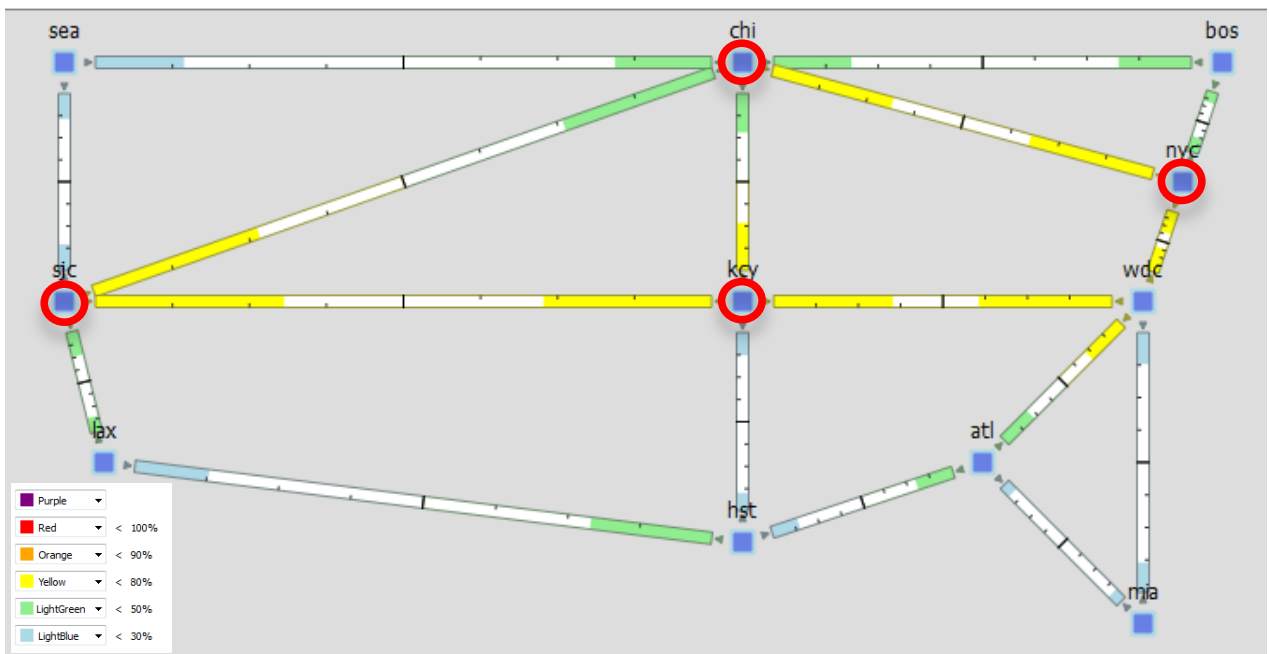

There are existing traffic demands on the network and the shading on the links indicates the current link utilisation.

Considering the example shown in Figure 2 the following process is used to determine which DC is the best choice to accommodate the new workload:

1. Requestor (e.g. orchestration system) sends a request for new workload to the Demand Engineering controller:

   i. Assume that the workload requires bidirectional demands of 100Mbps from every access site to the selected DC, i.e. as there are 11 access sites there will be 22 new demands in total.

ii. The required worst-case demand latency ($L_{max}$) is 25ms.

iii. The required worst-case path utilisation (T) is 95%, i.e. this is the maximum acceptable utilisation of the links on the path for the demands in order to ensure that the loss, latency and jitter SLAs required for the workload can be met.

iv. The optimisation policy is to *minimise worst-case path utilisation*

2. The controller adds the corresponding new demand(s) to its network model for each candidate data centre in turn as shown in Figure 3.    The scope of candidate DCs for that particular request may be constrained by the requestor or preconfigured in the controller; in this example the scope is unconstrained, i.e. all 4 DCs are considered candidates.

Figure 3.    Example Placement Decision (red lines show demands)

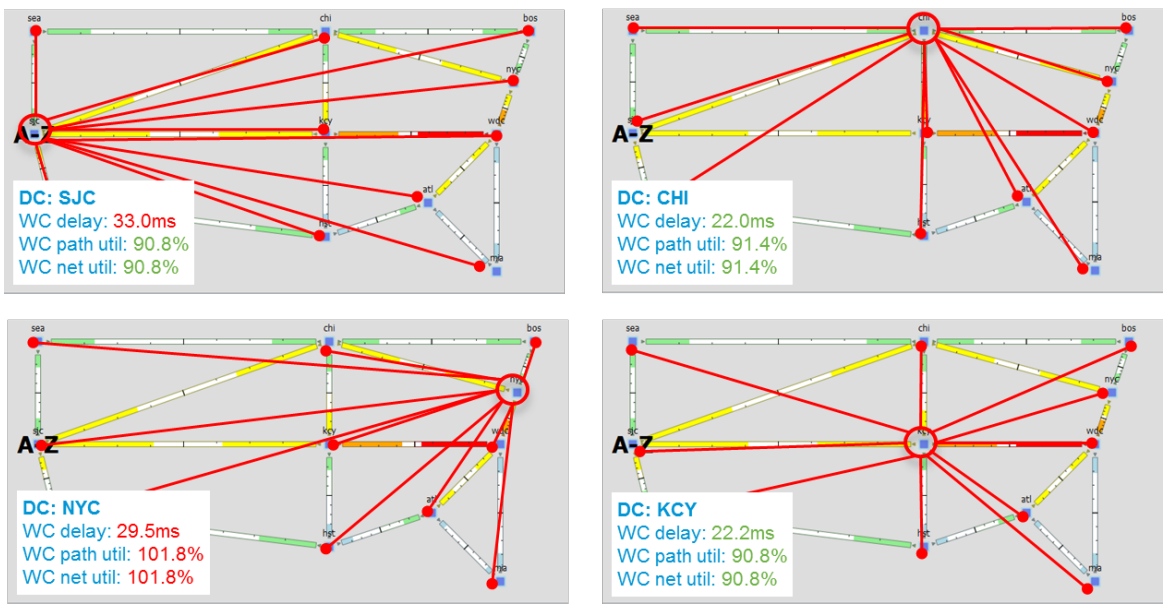

3. The controller uses simulation to determine:

i. The worst-case path latency for the requested demands, taking into account the required failure cases {no failures, links, nodes, SRLGs}. In this case siting the workload at the DC in either SJC or NYC results in exceeding the acceptable worst-case delay threshold, hence both SJC and NYC are discounted.

ii. The worst-case path utilisation for the requested demands, taking into account the required failure cases.    Siting the workload at the DC in NYC results in exceeding the acceptable worst-case path utilisation, hence NYC is discounted

iii. CHI and KCY can both support the requested demands with the required SLA, i.e. because worst-case delay, and worst-case path utilisation will be within bounds if the workload is sited at these DCs.    CHI and KCY are both considered feasible.

4. The controller applies the optimisation policy to decide which DC from the feasible list is the best to site the workload at.    It compares the corresponding worst-case path utilisation figures and chooses the DC that results in the lowest worst-case path utilisation.    In this case the DC at KCY is preferred because the worst-case path utilisation is lower than for CHI, i.e. siting the workload at KCY makes most efficient use of the available capacity.

5. The controller responds to the requestor with KCY as the preferred DC for the requested workload.

# 7. SIMULATION STUDY

We undertook a simulation study focussed on IAAS workload placement, using real network topologies from two service provider networks to determine the potential performance of demand engineering compared to other algorithms for determining workload placement.

One of the network topologies was from an international service provider interconnecting 11 locations, 7 of which provided connectivity to data centre resources; the network consisted of 10Gbps and 2.5Gbps links.    The other was from an in-country service provider interconnecting 5 locations, 2 of which provided connectivity to data centre resources; the network consisted of 80Gbps links.    We omit the actual topologies for reasons of confidentiality.

The following workload placement algorithms were compared:

- *Random DC selection* – this approach is network unaware; for each workload the serving DC is selected randomly.
- *Lowest latency DC selection* – where the DC is selected which has the lowest average path latency across the constituent demands of the workload.    This approach is aware of the network topology, but unaware of the network traffic.
- *Demand engineering DC selection* – where the DC is selected using demand engineering to minimise the worst-case path utilisation as described in this paper.    This approach is aware both of the network topology and the network traffic.

100 simulation iterations were run – each with over 500 randomly generated workloads:

- Each workload has a random selection of the 11 branch destinations
- The constituent demands for each workload were bidirectional and uniformly sized across the workload, i.e. each demand was the same bandwidth as the others in the workload
- The individual demand bandwidth was randomly selected for each workload from 50-500Mbps

Each simulation run was stopped when it was no longer possible to place demands without exceeding 100% path utilisation.

## Simulation Results

In order to compare the performance of the different algorithms we consider the aggregate demand throughput across all of the demands that were successfully placed across all simulation runs.

Figure 4 shows the detail of a single simulation run for the international service provider network. As can be seen, demand engineering ensures that the maximum path utilisation is consistently minimised, allowing more workload demands to be placed before 100% path utilisation is reached.

Figure 4.    Single simulation run results: In country service provider

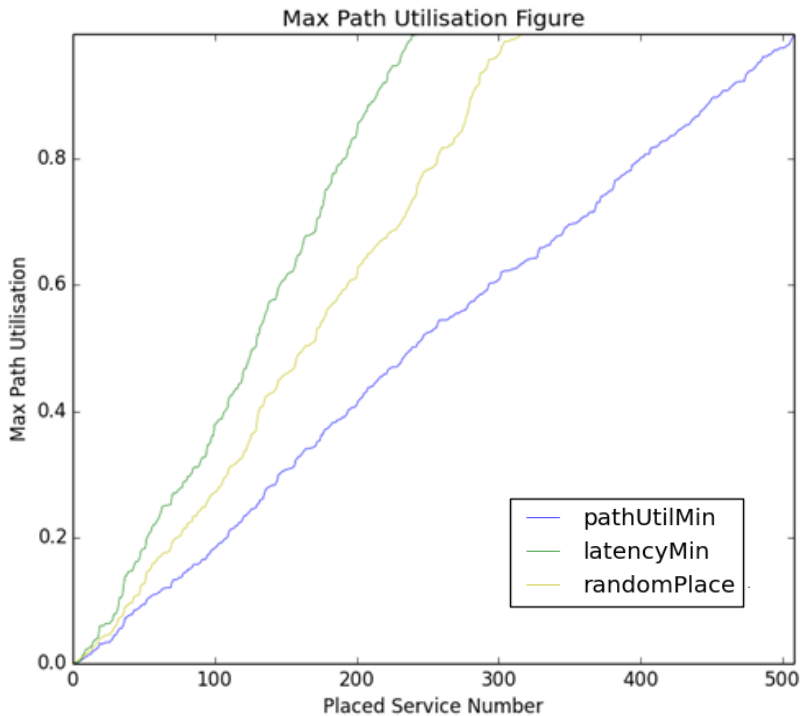

Figure 5 shows the results comparing the aggregate for the three workload placement algorithms for the international service provider topology across the 100 simulation runs. Figure 6 shows the corresponding results for the in country service provider topology.

Figure 5.    Aggregate results: International service provider

| Algorithm | Aggregate demand load for workloads placed | % of Demand Engineering Result |
|---|---|---|
| Minimise Latency | 485,408 | 48% |
| Random | 644,738 | 64% |
| Demand Engineering | 1,006,982 | 100% |

Figure 6.    Aggregate workload placement results: In country service provider

| Algorithm | Aggregate demand load for workloads placed | % of Demand Engineering Result |
|---|---|---|
| Minimise Latency | 5,682,688 | 86% |
| Random | 5,230,336 | 79% |
| Demand Engineering | 6,607,616 | 100% |

The International Service Provider results show that demand engineering provides an effective bandwidth gain of 56% over random placement and 108% over lowest latency placement.    This network topology has little symmetry and the 2.5Gbps links were the dominating constraint; demand engineering is able to distribute the workloads taking these into account which results in a significant gain over network unaware approaches.

For the in-country provider demand engineering provides an effective bandwidth gain of 26% over random placement and 16% over lowest latency placement.    In this case the network topology and capacity are symmetrical and benefit from equal cost multipath routes between all locations, which results in better balancing of load by default.    Even though this is the case, an uneven distribution of data centre resources on the topology results in a potential mismatch between the traffic demands and installed capacity such that demand engineering is still able to provide a tangible gain.

Simulation for other applications and services is a potential area for future study.

# 8. DEMAND ENGINEERING COMPARED TO TRAFFIC ENGINEERING

We compare and contrast Demand Engineering and Traffic Engineering:

- *OSI Layer.*    Demand Engineering and TE operate at different layers; TE is a network-layer capability whereas Demand Engineering has network-awareness but directly influences application-layer behaviour.
- *Implicit / Explicit Paths.*    TE explicitly determines the paths that demands take through the network between their predefined traffic sources and destinations.    In contrast, Demand engineering directly influences the location of traffic sources and destinations, i.e. which implicitly impacts the paths that the application or services traffic demands take through the network.    As a consequence, Traffic engineering can implement arbitrary network routing policies whereas demand engineering cannot.
- *Admission control.*    Demand engineering implicitly provides an admission control capability.    TE could be augmented by admission control mechanisms such as RSVP [8], however, in practise RSVP has failed to gain widespread use.

- *Optimisation*:
    - *Efficacy.*   The results of the simulation study indicate that in terms of network optimisation, the effective bandwidth gain from Demand Engineering is comparable with the results achieved with traffic engineering [3].
    - *Periodic / Continuous.*   Traffic Engineering requires periodic re-optimisation to deal with changes in the network topology and traffic matrix.   Demand engineering, however, is a proactive and transactional process that is applied at the time of service instantiation; where there is service churn this can provide a continuous optimisation, i.e. which therefore does not require reoptimisation.
- *Applicability.*   As a network-layer capability TE has no dependencies on high layers and can be applied to all traffic.   Demand engineering is applicable to applications and services where there is a choice as to where the application endpoints may be located; we estimate applicability to ~90% of the traffic generated by applications and services today.

Although we have compared their differences, because they operate at different layers Demand Engineering and TE can be complimentary, and may be deployed independently or in concert.   It is possible to use a single network controller to be used for TE and Demand Engineering simultaneously for maximum possible bandwidth gain.

# 9. CONCLUSION

In this paper we introduce Demand engineering, which is a new approach to traffic management which leverages a network and traffic-aware controller to determine the "best" location to site or locate an application or service instance, where best is determined using an understanding of the network topology and traffic and is defined in terms of both meeting the SLA requirements and making most effective use of the network capacity.

Demand engineering maximises the demands that can be serviced by placing demands where there are available network resources.   The effective bandwidth gain that demand engineering provides will be dependent on the topology and traffic matrix (as is the case for traffic engineering); our studies show gains of between 16% and 56% over network unaware placement approaches for the different networks studied.

Demand engineering provides the benefits of both admission control and traffic engineering without suffering the issues of previous approaches that have limited their respective deployment.   This stands to benefit both network service providers and service consumers through making more optimal use of network capacity, hence minimising cost, and assuring more deterministic SLAs with improved quality of experience.

Following this study, we have built a working implementation of demand engineering in Cisco WAN Automation Engine (WAE) [12].

# 10. FUTURE WORK

The work described in this paper focusses on network-aware demand placement. With the deployment cloud services and Network Function Virtualisation (NFV), there are choices as to where each service or workload may be placed; in these cases, a workload requires both network and data centre resources. The concept of Demand Engineering can be extended to be both network and data centre aware, i.e. to address the placement of workloads taking into account both network and compute, storage, memory resources etc. We highlight this as an area of interest for further study.